 \documentclass[aps,prb,twocolumn,showpacs]{revtex4-1}

\usepackage[english]{babel}
\usepackage[]{graphicx}
\usepackage{amsmath}
\usepackage{times}
\usepackage{bm}
\usepackage{units}
\usepackage[normalem]{ulem}
\usepackage{color}

\newcommand{\comment}[1]{}

\newcommand{\braket}[1]{\langle #1\rangle}

\begin{document}

\title{Microscopic description of intraband absorption in graphene: the occurrence of transient negative differential transmission}

\author{Faris Kadi$^{1}$}
\email[]{faris.kadi@mailbox.tu-berlin.de}
\author{Torben Winzer$^{1,2}$}	
\author{Ermin Malic$^{1}$}
\author{Andreas Knorr$^{1}$}

\affiliation{$^{1}$Institut f\"ur Theoretische Physik, Nichtlineare Optik und Quantenelektronik, Technische Universit\"at Berlin,
  Hardenbergstr. 36, 10623 Berlin, Germany}
	
\affiliation{$^{2}$Department of Materials Science and Engineering, Yonsei University,
Seoul 120-749, Korea}	
	
\author{F. G\"ottfert$^{3}$, M. Mittendorff$^{3}$, S. Winnerl$^{3}$}
\author{M. Helm$^{3}$}
\affiliation{$^{3}$Helmholtz-Zentrum Dresden-Rossendorf, PO Box 510119, D-01314 Dresden, Germany }

%%%%%%%%%%%%%%%%%%%%%%%%%%%%%%%%%%%%%%%%%%%%%%%%%%%%%%%%%%%%% 
\begin{abstract}
We present a microscopic explanation of the controversially discussed transient negative differential transmission observed in degenerate optical pump-probe 
measurements in graphene.
Our approach is based on the density matrix formalism allowing a time- and momentum-resolved study of carrier-light, carrier-carrier, and carrier-phonon 
interaction on microscopic footing. 
We show that phonon-assisted optical intraband transitions give rise to transient absorption in the optically excited hot carrier system counteracting 
pure absorption bleaching of interband transitions. 
While interband transition bleaching is relevant in the first hundreds of fs after the excitation, intraband absorption sets in at later times.
 In particular, in the low excitation regime,
 these intraband absorption processes prevail over the absorption bleaching resulting in a zero-crossing of the differential transmission. Our findings are in good agreement with recent experimental pump-probe studies.
\end{abstract}

\maketitle
%%%%%%%%%%%%%%%%%%%%%%%%%%%%%%%%%%%%%%%%%%%%%%%%%%%%%%%%%%%%%
%\onecolumn

The ultrafast carrier relaxation dynamics in optically excited graphene has been intensively studied.\cite{sun08,dawlaty08,plochocka09,wang10,zou10,breusing11,winnerl11,heinz13-3,obraztsov11,winnerl13,winzer12-1}
Typically, the carrier relaxation has been accessed via high-resolution pump-probe experiments. \cite{sun08,breusing11,winnerl11,strait11,obraztsov11}
Common to all studies is a bi-exponential decay of the pump-induced differential transmission (DT) spectrum.
The fast decay component in the range of few tens of femtoseconds is assigned to an ultrafast Coulomb-dominated carrier redistribution towards a hot Fermi-Dirac distribution, whereas the slower decay component in the range of a picosecond reflects the equilibration between the electron and the phonon system.\cite{breusing11}
However, while some of the studies report on a purely positive transient DT spectrum,\cite{dawlaty08,wang10,kumar09} others exhibit a zero-crossing after the initial decay.\cite{newson09,plochocka09,sun08,sun10-1,winnerl11,shang10,breusing11} 
The second decay component 
characterizes the recovering of the negative transient DT signal. 
There is a number of possible underlying mechanisms for the DT-zero crossing.\cite{winnerl11,winnerl13, erminbuch,newson09,plochocka09,sun08,sun10-1,shang10,breusing11}
For some of the experiments, the negative DT can be clearly traced back to the predominant intraband absorption. This is the case, when the photon energy is smaller than twice the value of the Fermi 
energy.\cite{sun08,winnerl11,sun10-1} However, for all other cases, where negative DT components occur for photon energies much larger than the Fermi energy, the underlying mechanism has been controversially discussed in literature.
In particular, pump-induced bandstructure renormalization for transient \cite{breusing11} or in the long time limit \cite{binder13}  at the $M$-point as well as intraband absorption\cite{heinz13-3} have been suggested as explanations for the observed negative DT signals.
\begin{figure}[t!]
  \begin{center}
\includegraphics[width=0.51\linewidth]{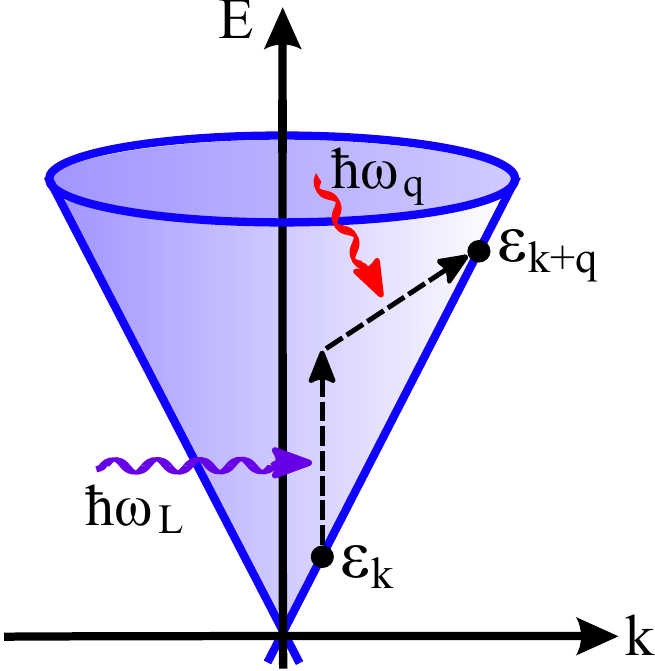}
  \end{center}
  \caption{Illustration of the phonon-assisted intraband absorption within the linear band structure of graphene.
  Applying a pulse with the energy $\hbar\omega_L$ lifts electrons within the (conduction) band into an energetically 
  higher electronic state. To conserve energy and momentum, this intraband absorption is assisted by an interaction with a phonon
  characterized by the energy $\hbar\omega_{\bf q}$ and the momentum $\bf q$. 
  } \label{fig1} 
\end{figure}
In this Letter, we present a microscopic explanation for the occurrence of transient negative differential transmission in graphene. Based on the density matrix formalism, we investigate the 
 detailed interplay of intra- and interband absorption processes on the transient DT in graphene.
The absorptive electronic intraband transitions are assisted by emission or absorption of a phonon with an adequate momentum necessary to fulfill the momentum conservation, cf. Fig. \ref{fig1}.  
\textit{Intraband transitions are shown to lead to an enhanced absorption giving rise to the experimentally observed zero-crossing from positive to negative DT signals.} Note that
these intraband absorption processes are entirely related to photogenerated carriers far from equilibrium as the calculations are performed for intrinsic graphene.
Existing studies taking into account intraband absorption based on the Kubo formalism \cite{wu13} and the Drude model \cite{heinz13-3,choi09,mikhailov07,Falkovsky07} are typically stationary and rely on phenomenological scattering rates. Our 
description goes significantly beyond these studies as we include time- and momentum-dependent phonon-assisted intraband absorption in a fully optical excited non-equilibrium situation on the same microscopic footing as the direct intra- and interband 
carrier-carrier and carrier-phonon scattering relaxation channels.
In particular, we show that the parameter-free calculated transient ratio of inter- and intraband processes can explain many recent pump-probe experiments in graphene showing an interplay of positive and negative transmission. \\

The absorption coefficient $\alpha(\omega)$ of graphene is determined by the imaginary
part of the optical susceptibility $\chi(\omega)=j(\omega) / (\epsilon_0 \omega^2 A(\omega))$ with the current density $j(\omega)$, the
vector potential $A(\omega)$, and the dielectric permittivity $\epsilon_0$.\cite{erminbuch} The current  reads
\begin{equation}
 \label{current}
{\bf j}(t)=B\sum_{\bf k} \left[2{\bf M}_{\bf k}^{vc} \Im[p_{{\bf k}}]  -i\sum_{\lambda}\left({\bf M}_{\bf k}^{\lambda\lambda} \rho_{\bf k}^\lambda + \frac{e_0}{\hbar} {\bf A}(t)\rho_{\bf k}^\lambda \right) \right] 
\end{equation}
with the constant $B=2e_0 \hbar / (m_0L^2)$, where $e_0$ denotes the elementary charge, $m_0$ the free electron mass and $L^2$ the structure area that cancels out after performing the sum over $\bf k$. The current contains an interband (first term) and an intraband contribution (second and third term). The coupling to the light is described by the interband or intraband optical matrix element ${\bf M}_{\bf k}^{\lambda \lambda'}$, respectively.\cite{malic11-1} 
Furthermore, the current depends on the carrier occupation probability $\rho_{{\bf k}}^{\lambda}$ with
the band index $\lambda$ and the momentum $\bf k$ and
the microscopic polarization $p_{\bf k}$ that is a measure for the optical transition probability.\cite{malic11-1} 
These two microscopic quantities are determined by solving graphene Bloch equations. \cite{malic11-1}
It can be shown that the third term of the current ${\bf j}(t)$ in Eq. (\ref{current}) cancels with the real part of interband term resulting in a purely imaginary current.\cite{sipe93}
Finally, the absorption coefficient can be written as 
\begin{align}
\alpha(\omega)= \frac{B}{\omega c_0 \varepsilon_0}\sum_{\bf k}\Im\left[\frac{2{\bf M}_{\bf k}^{vc} 
\Im[p_{{\bf k}}(\omega)]}{A(\omega)}  -i\sum_{\lambda}\frac{{\bf M}_{\bf k}^{\lambda\lambda} \rho_{\bf k}^\lambda(\omega)}{A(\omega)}\right] \label{abs}  {\rm ,} 
\end{align}
with %$B=2e_0\hbar / (c_0 \varepsilon_0  m_0L^2)$ and 
the velocity of light $c_0$. 
The interband contribution is proportional to ${\bf M}_{\bf k}^{vc}$ and to the microscopic polarization $p_{{\bf k}}(\omega)$ leading to the well-known absorption spectrum of graphene with a constant value in the visible spectral range.\cite{erminbuch,mak11,stauber08,nair08,heinz08,geim07}
The phonon-assisted intraband processes influence the spectra via the carrier occupation probability $\rho_{\bf k}^\lambda(\omega)$ weighted by the optical intraband matrix element ${\bf M}_{\bf k}^{\lambda\lambda}$.

Since the main goal of our work is the description of the non-equilibrium regime, we focus on the differential transmission spectrum that is determined by $\Delta T / T_0(\tau,\omega) \propto \alpha^{(t)}(\omega)-\alpha^{(p,t)}(\omega,\tau)$
with the absorption coefficient including both the pump and the probe (test) 
pulse $\alpha^{(p,t)}(\omega,\tau)$ and only the test pulse 
$\alpha^{(t)}(\omega)$.\cite{mukamel99}
Assuming an isotropic carrier distribution in the momentum space after several fs,\cite{malic12} the transient DT reads  
\begin{align}\label{dtseq}
&\Delta T / T_0(\tau,\omega) \propto C_{k_\omega}[\Delta \rho^{(p,t)}_{k_\omega}(\tau)-\Delta \rho^{(t)}_{k_\omega}(\tau)] \\ \nonumber 
 &+ \frac{i}{L^2 \omega A^{(t)}_0}\sum_{\lambda{\bf k}}{\bf M}_{\bf k}^{\lambda\lambda} \Im[\rho^{\lambda,(p,t)}_{\bf k}(\omega,\tau)- \rho^{\lambda,(t)}_{\bf k}(\omega,\tau)]  {\rm ,}
\end{align}
with $C_{k_\omega}=|{\bf M}_{k_\omega}^{vc}|^2 e_0 \hbar^2 / (8\pi m_0 v_F^2)$, $\Delta\rho_{k} = \rho^c_{k} - \rho^v_{k}$, the 
amplitude of the vector potential $A_0$ and the excitation momentum $k_\omega=\hbar \omega_L/2v_F$ with the pulse frequency $\omega_L$ and the carrier velocity close to the Dirac point $v_F$. 
The first line of Eq. (\ref{dtseq}) accounts for interband transitions, which reflects the absorption bleaching due to the additional carriers lifted from the valence into the conduction band by the pump pulse. 
As a result, the transient DT is expected to show a positive peak during the excitation. The second line describes the phonon-assisted intraband absorption processes that are induced by the probe pulse. 
Since these processes lead to an increased absorption after the excitation, we expect also a negative contribution to the transient DT signal.
\\

To determine the interplay of these two contributions, we derive equations of motion for the carrier occupation probability $\rho^\lambda_{\bf k}$ and 
the microscopic polarization $p_{\bf k}$. To exploit the Heisenberg equation, we need the many-particle Hamilton operator $H = H_0 + H_{c,f} + H_{c,p} + H_{c,c}$ with $H_0$ 
denoting the interaction-free carrier and phonon part, 
$H_{c,f}$ the carrier-field coupling, 
$H_{c,p}$ the carrier-phonon interaction, and
$H_{c,c}$ the carrier-carrier interaction.\cite{erminbuch,malic11-1}  
In this letter, we focus on the dynamics of the carrier occupation  driven by the electron-phonon contribution $\dot{\rho}_{{\bf k}}^{\lambda}|_{H_{c,p}}$, which determines the phonon-assisted intraband absorption processes.\cite{kira03}
Other contributions and the equation for the microscopic polarization can be found in Ref. \onlinecite{malic11-1}. Here, we discuss 
only the carrier-phonon interaction with $H_{c,p}=\sum_{{\bf l}_1{\bf l}_2}\sum_{{\bf u}}(g_{{\bf u}}^{{\bf l}_1{\bf l}_2}a_{{\bf l}_1}^{\dagger} a_{{\bf l}_2}^{\phantom{\dagger}}b_{{\bf u}}+g_{{\bf u}}^{{\bf l}_1{\bf l}_2*}a_{{\bf l}_2}^{\dagger} a_{{\bf l}_1}^{\phantom{\dagger}}b^\dagger_{-{\bf u}}),$ 
where $a_{{\bf l}}^{\dagger}$ and $a_{{\bf l}}$ create and annihilate an electron in the state ${\bf l}=(\lambda, {\bf k})$, respectively. Furthermore, we introduce the bosonic operators $b^\dagger_{{\bf u}}$ and
$b_{{\bf u}}$, which create and annihilate a phonon in the state ${\bf u}=(j, {\bf q})$ with the phonon momentum ${\bf q}$ and the different optical and acoustic phonon modes $j$.
The strength of the carrier-phonon interaction is determined by the corresponding electron-phonon coupling elements $g_{{\bf u}}^{{\bf l}_1{\bf l}_2}$.\cite{hwang08-1,piscanec04}
The phonon dispersion $\hbar\omega_{\bf q}^j$ is assumed to be constant for optical modes close to the $\Gamma$- and $K$-point [\onlinecite{piscanec04}] and
linear for acoustic modes [\onlinecite{hwang08-1}] close to the $\Gamma$-point.
Applying the Heisenberg equation of motion, we obtain 
\begin{align}
 \nonumber  \dot{\rho}_{{\bf k}}^{\lambda}|_{H_{c,p}} =\frac{1}{i\hbar}&\sum_{\lambda'}\sum_{j {\bf q}}\Bigl( g_{{\bf k},{\bf q}}^{\lambda \lambda' j}S^{\lambda \lambda' j}_{{\bf k},{\bf q}}-g_{{\bf k}+{\bf q},{\bf q}}^{\lambda' \lambda j}S_{{\bf k}+{\bf q},{\bf q}}^{\lambda' \lambda j}\\ 
 & \ \ \ \ \ \ \ \ \ - g_{{\bf k},{\bf q}}^{\lambda \lambda' j*}T^{\lambda' \lambda j}_{{\bf k},{\bf q}}  +g_{{\bf k}+{\bf q},{\bf q}}^{\lambda' \lambda j*}T_{{\bf k}+{\bf q},{\bf q}}^{\lambda \lambda' j}  \Bigr)\rm{,} \label{phononyeah}
\end{align}
where $S^{\lambda \lambda' j}_{{\bf k},{\bf q}}=\braket{a^{\lambda \dagger}_{\bf k}a^{\lambda'}_{{\bf k}-{\bf q}}b_{\bf q}^j}$ and 
$T^{\lambda \lambda' j}_{{\bf k},{\bf q}}=\braket{a^{\lambda \dagger}_{{\bf k}-{\bf q}}a^{\lambda'}_{{\bf k}}b_{\bf q}^{j \dagger}}$
are phonon-assisted electron densities and transitions.\cite{schilp94} 
To obtain a closed system of differential equations, we derive additional equations for the phonon-assisted quantities.\cite{schilp94,rossi94} Exemplary, we show the equation for $S^{\lambda_1 \lambda_2 j}_{{\bf k},{\bf q}}$ reading
\begin{align}
 \nonumber &\dot{S}^{\lambda_1 \lambda_2 j}_{{\bf k},{\bf q}} = \frac i \hbar \Delta\varepsilon_{{\bf k},{\bf q}}^{\lambda_1 \lambda_2} S^{\lambda_1 \lambda_2 j}_{{\bf k},{\bf q}} + \frac i \hbar \sum_{\lambda_3 \lambda_4}g_{{\bf k},{\bf q}}^{\lambda_3 \lambda_4 j*} Q_{{{\bf k},{\bf q}}}^{\lambda_1 \lambda_3 \lambda_4 \lambda_2 j} \\
& \ +\frac{\hbar e_0}{m_0} \sum_{\lambda'} \left[ S^{\lambda_1 \lambda' j}_{{\bf k},{\bf q}} {\bf M}^{\lambda_2 \lambda'}_{{\bf k}-{\bf q}}  - S^{\lambda' \lambda_2 j}_{{\bf k},{\bf q}} {\bf M}^{\lambda' \lambda_1}_{{\bf k}}  \right]{\bf A}(t) {\rm ,} \label{SSS}
\end{align}
with the energy $\Delta\varepsilon_{{\bf k},{\bf q}}^{\lambda_1\lambda_2} =\varepsilon_{\bf k}^{\lambda_1}-\varepsilon_{{\bf k}-{\bf q}}^{\lambda_2}-\hbar\omega_{\bf q}^j$ and 
the scattering kernel $Q_{{\bf k},{\bf q}}^{\lambda_1 \lambda_2 \lambda_3 \lambda_4 j}=(\delta_{\lambda_1 \lambda_2}-\sigma_{\bf k}^{\lambda_1 \lambda_2})\sigma^{\lambda_3 \lambda_4}_{{\bf k}-{\bf q}} n_{\bf q}^j 
 - \sigma_{\bf k}^{\lambda_1 \lambda_2}(\delta_{\lambda_3 \lambda_4}-\sigma_{{\bf k}-{\bf q}}^{\lambda_3 \lambda_4})(n_{\bf q}^j+1)$, where 
$\sigma_{\bf k}^{\lambda \lambda'}=\braket{a^{\lambda \dagger}_{\bf k}a^{\lambda'}_{\bf k}}$ and 
$n^j_{\bf q}$ is the phonon occupation. 
The corresponding equation for $T_{{\bf k},{\bf q}}^{\lambda \lambda' j}$ can be  derived in analogy. \cite{malic11-1, erminbuch}

The last term in Eq. (\ref{SSS}) is of crucial interest for us, since it describes phonon-assisted optical transitions. They are driven by both carrier-light interaction (${\bf M}_{\bf k}$) and implicitly also by carrier-phonon matrix elements appearing in the dynamics of the phonon-assisted quantity $S^{\lambda_1 \lambda_2 j}_{{\bf k},{\bf q}}$.
Our goal is the description of the phonon-assisted intraband absorption processes affecting the weak probe pulse applied after a much stronger pump pulse that creates a non-equilibrium carrier distribution.
Since the experimentally observed negative transient DT spectrum occurs clearly after the application of the optical excitation, we study the interaction of the test pulse with the carrier occupation resulting from the pump-induced
optical dynamics.\cite{breusing11,winnerl13}
To be able to solve Eq. (\ref{SSS}) analytically, we chose a perturbative approach assuming that \cite{kira03}
\begin{align}
S^{\lambda_1 \lambda_2 j}_{{\bf k},{\bf q}}=S^{\lambda_1 \lambda_2 j}_{{\bf k},{\bf q}(0)}+\epsilon S^{\lambda_1 \lambda_2 j}_{{\bf k},{\bf q}(1)} {\rm ,} \label{SSS2}
\end{align}
where $\epsilon$ represents a small perturbation corresponding to the  amplitude of the vector potential ${\bf A}_{\text{test}}(t) = {\bf A}_0 e^{-\frac{t^2}{2\sigma^2}}\cos(\omega_L t)$ describing the probe pulse with $\sigma$ as the pulse duration. 
Within the well tested Markov approximation,\cite{malic11-1} the zeroth and the first order of $S^{\lambda_1 \lambda_2 j}_{{\bf k},{\bf q}}$, Eq. (\ref{SSS2}), are obtained analytically reading:
\begin{align}
&S^{\lambda_1 \lambda_2 j}_{{\bf k},{\bf q}(0)} = i\pi \sum_{\lambda_3 \lambda_4}g_{{\bf k},{\bf q}}^{\lambda_3 \lambda_4 j*} 
Q_{{{\bf k},{\bf q}}}^{\lambda_1 \lambda_3 \lambda_4 \lambda_2 j}\delta (\Delta \varepsilon^{\lambda_3 \lambda_4 }_{{\bf k},{\bf q}}) {\rm ,} \label{s_0} \\
\nonumber &S^{\lambda_1 \lambda_2 j}_{{\bf k},{\bf q}(1)} = i\pi \sum_{\lambda_3 \lambda_4}g_{{\bf k},{\bf q}}^{\lambda_3 \lambda_4 j*} 
Q_{{{\bf k},{\bf q}}}^{\lambda_1 \lambda_3 \lambda_4 \lambda_2 j}\frac{  \frac{i e_0}{m_0} \Delta {\bf M}^{\lambda \lambda'}_{{\bf k},{\bf q}}e^{-\frac{t^2}{2\sigma^2}} 
    }{2 \Delta \varepsilon^{\lambda_1 \lambda_2 }_{{\bf k},{\bf q}} } \\
 &[e^{-i\omega_Lt}\delta(\Delta \varepsilon^{\lambda_3 \lambda_4 }_{{\bf k},{\bf q}}+\hbar\omega_L)+e^{i\omega_Lt}\delta(\Delta 
 \varepsilon^{\lambda_3 \lambda_4 }_{{\bf k},{\bf q}}-\hbar\omega_L) ] \label{s_1} , 
\end{align}
with $\Delta {\bf M}^{\lambda \lambda'}_{{\bf k},{\bf q}}=  ({\bf M}_{{\bf k}-{\bf q}}^{\lambda' \lambda'} - {\bf M}_{{\bf k}}^{\lambda \lambda } )$.
All terms proportional to the off-diagonal matrix element ${\bf M}^{vc}_{\bf k}$ turn out to cancel in the equation of $\rho_{{\bf k}}^\lambda$.
The first order $S^{\lambda_1 \lambda_2 j}_{{\bf k},{\bf q}(1)}$ describes the phonon-assisted intraband absorption, while the zeroth order $S^{\lambda_1 \lambda_2 j}_{{\bf k},{\bf q}(0)}$
 yields  the standard Boltzmann-like equation containing phonon-assisted intra- and interband scattering processes. 
 
Using Eq. (\ref{s_0}) and (\ref{s_1}) the equation of motion for the carrier occupation $\rho_{\bf k}^\lambda$ can be written as
\begin{align}
\label{rho_cp} \dot{\rho}_{{\bf k}}^{\lambda}|_{H_{c,p}+H_{c,c}}= \tilde{\Gamma}^{\text{in}}_{\lambda,{\bf k}}\left(1-\rho_{{\bf k}}^{\lambda}\right)-\tilde{\Gamma}^{\text{out}}_{\lambda,{\bf k}}\rho_{{\bf k}}^{\lambda} {\rm ,}
\end{align}
 with $\tilde{\Gamma}^{\text{in,out}}_{\lambda,{\bf k}} = \Gamma^{\text{in,out}}_{\lambda,{\bf k}}+\Gamma^{\text{in,out,A}}_{\lambda,{\bf k}}$.
Here, $\Gamma^{\text{in,out}}_{\lambda,{\bf k}}$ describes the time- and momentum-dependent scattering rates including the Coulomb- and phonon-assisted contributions, which are in detail discussed in Ref. \onlinecite{malic11-1}.
The field-assisted second order contribution $\Gamma^{\text{in,A}}_{\lambda,{\bf k}}$, responsible for the intraband absorption, reads:
\begin{align}
&\Gamma^{\text{in,A}}_{\lambda,{\bf k}}= 2\pi\sum_{\lambda' j{\bf q}} \left[|g^{\lambda' \lambda j}_{{\bf k},{\bf q}}|^2 \left(\frac{  
\tilde{{\bf M}}^{\lambda'\lambda}_{{\bf k},{\bf q}}  }
{\Delta \varepsilon^{\lambda \lambda'}_{{\bf k},{\bf q}}}\delta(\Delta \varepsilon^{\lambda \lambda'}_{{\bf k},{\bf q}}\pm\hbar\omega_L) \right) \rho_{{\bf k}-{\bf q}}^{\lambda'} n_{\bf q}^j \right. \\\nonumber
  &\left.+|g^{\lambda' \lambda j}_{{\bf k}+{\bf q},{\bf q}}|^2 \left(\frac{  \tilde{{\bf M}}^{\lambda\lambda'}_{{\bf k}+{\bf q},{\bf q}} }
	{\Delta \varepsilon^{\lambda' \lambda}_{{\bf k}+{\bf q},{\bf q}}}\delta(\Delta \varepsilon^{\lambda' \lambda}_{{\bf k}+{\bf q},{\bf q}}\pm\hbar\omega_L) \right) \rho_{{\bf k}+{\bf q}}^{\lambda'} (n_{\bf q}^j+1) \right]
\end{align}
with $\tilde{{\bf M}}_{{\bf k},{\bf q}}^{\lambda \lambda' } = i e_0 / m_0 ({\bf M}_{{\bf k}-{\bf q}}^{\lambda' \lambda'} - 
{\bf M}_{{\bf k}}^{\lambda \lambda } ){\bf A}(t)$.
It accounts for optically driven intraband absorption processes that are assisted by phonon absorption or emission fulfilling the momentum and energy conservation.
The term $\Gamma^{\text{out,A}}_{\lambda,{\bf k}}$ can be obtained analogously by replacing $\rho \leftrightarrow \rho-1$ and 
$n \leftrightarrow n+1$.
\begin{figure}[t!]
  \begin{center}
\includegraphics[width=1.00\linewidth]{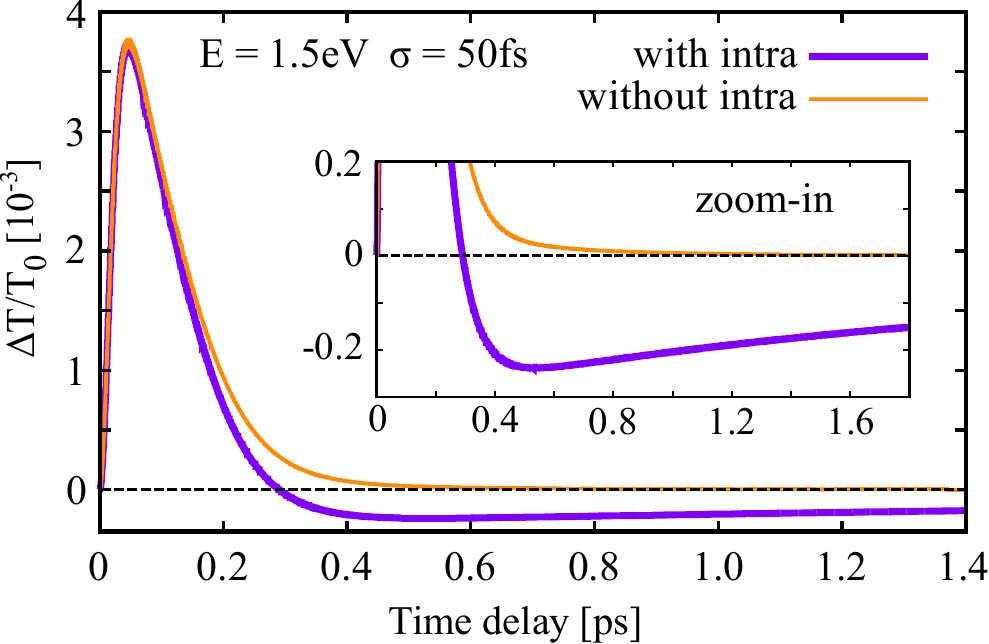}
  \end{center}
  \caption{Transient differential transmission spectrum as a function of the delay time $\tau$ between the pump and the probe pulse illustrating the impact of phonon-assisted intraband processes, which account for a zero-crossing.
	}\label{fig2} 
\end{figure} 
Basically, these terms have a similar form as the scattering rates $\Gamma^{\text{in,out}}_{\lambda,{\bf k}}$. Since they are proportional to the optical matrix elements $\tilde{{\bf M}}^{\lambda\lambda'}_{{\bf k},{\bf q}}$, we expect 
them to have a contribution to the DT.\\
  Similarly to Eq. (\ref{SSS}),   Eq. (\ref{rho_cp}) can be further simplified by using the ansatz $\rho_{{\bf k}}^{\lambda} (t) = \rho_{{\bf k}}^{\lambda,0}(t) + \delta\rho_{{\bf k}}^{\lambda} (t)$. Here, 
$\rho_{{\bf k}}^{\lambda,0}(t)$ contains the pump-induced 
occupation resulting from the Coulomb- and phonon-assisted scattering rates $\Gamma^{\text{in,out}}_{\lambda,{\bf k}}$,\cite{erminbuch} while 
$\delta\rho_{{\bf k}}^{\lambda}(t)$ 
describes a small perturbation induced by the probe pulse. The latter is in particularly driven by phonon-assisted intraband absorption processes. 
Assuming a weak $\delta$-shaped probe pulse ${\bf A}= {\bf A}_0 \delta(t-\tau)$, 
the Fourier transform for the perturbation yields
\begin{align}
\label{drho}
\delta \rho_{{\bf k}}^{\lambda}(\omega,\tau) = \frac{\Gamma^{\text{in,A}}_{\lambda,{\bf k}}(\omega,\tau)\left(1-\rho_{{\bf k}}^{\lambda,0}(\tau)\right) - 
\Gamma^{\text{out,A}}_{\lambda,{\bf k}}(\omega,\tau)\rho_{{\bf k}}^{\lambda,0}(\tau)}{i\omega + \tilde{\gamma}}
\end{align}
where the small damping  $\tilde{\gamma}=\Gamma^{\text{in}}_{\lambda,{\bf k}}+\Gamma^{\text{out}}_{\lambda,{\bf k}}$ is assumed to be constant. 
Note that this quantity has only a marginal influence on the impact of intraband absorption processes since all different
wavenumber contributions have to be summed up, cf. Eq. (\ref{abs}). In the expression of
$\delta \rho_{{\bf k}}^{\lambda}(\omega,\tau)$, we neglect the non-linear terms proportional to ${\bf A}(t)$ assuming a weak probe pulse.\\

Inserting Eq. (\ref{drho}) into Eq. (\ref{dtseq}), we have all ingredients at hand to evaluate the impact of phonon-assisted intraband 
absorption processes on the transient differential transmission (Fig. \ref{fig2}).
As initial condition, we assume a Fermi distribution  for the electron population $\rho_{\bf k}$ and a Bose-Einstein distribution for the phonon occupations $n^j_{\bf q}$ at room temperature. 
First, we analyze the impact of the excitation strength on the transient DT spectrum. Figure \ref{fig2} shows the calculated degenerate transient DT spectrum $\Delta T / T_{0}(\tau)$ excited by a $50$ fs pulse
with a pump fluence of $8$ $\mu$J/cm$^2$ at the photon energy $E = 1.5$ eV.
The abrupt increase of the signal reflects the ultrafast optical injection of non-equilibrium carriers around the photon energy resulting in an absorption bleaching. Afterwards, the signal decreases on a 
fs timescale due to the carrier-carrier (thermalization of the non-equilibrium electrons) and carrier-phonon (cooling and recombination of the thermalized electrons) scattering.\cite{breusing11} 
Taking into account only interband absorption (Fig. \ref{fig2}, thin line), the transient DT signal remains positive during the entire dynamics. 
However, after  including the phonon-assisted intraband absorption, 
the transient exhibits a distinct zero-crossing after approximately $310$ fs followed by a subsequent recovery towards zero on a ps timescale.
This observation reflects well the experimental observation, as will be discussed below. 

For both theory curves in Fig. \ref{fig2}, we find a bi-exponential decay characterized by a fast time constant $\tau_1=90$ fs and a slow 
component $\tau^{\text{inter}}_2=0.6$ ps (positive DT, thin orange line) and $\tau^{\text{inter+intra}}_2=2.8$ ps (negative DT, thick violet line) depending on 
whether intraband processes are included or not. The intraband absorption does not change the first decay rate, however, the second decay is considerably slowed down in the presence of intraband absorption. 
 The decay $\tau^{\text{inter}}_2$ is determined by the electron cooling due to electron-phonon interaction, whereas the decay 
$\tau^{\text{inter+intra}}_2$ in the presence of intraband absorption has a different origin: Since intraband absorption occurs always for finite carrier population $\rho_{\bf k}^\lambda$, $\tau^{\text{inter+intra}}_2$ is 
limited by carrier recombination processes that take place on a slower time scale (several ps) than the carrier cooling (one ps). 

\begin{figure}[t!]
  \begin{center}
\includegraphics[width=0.97\linewidth]{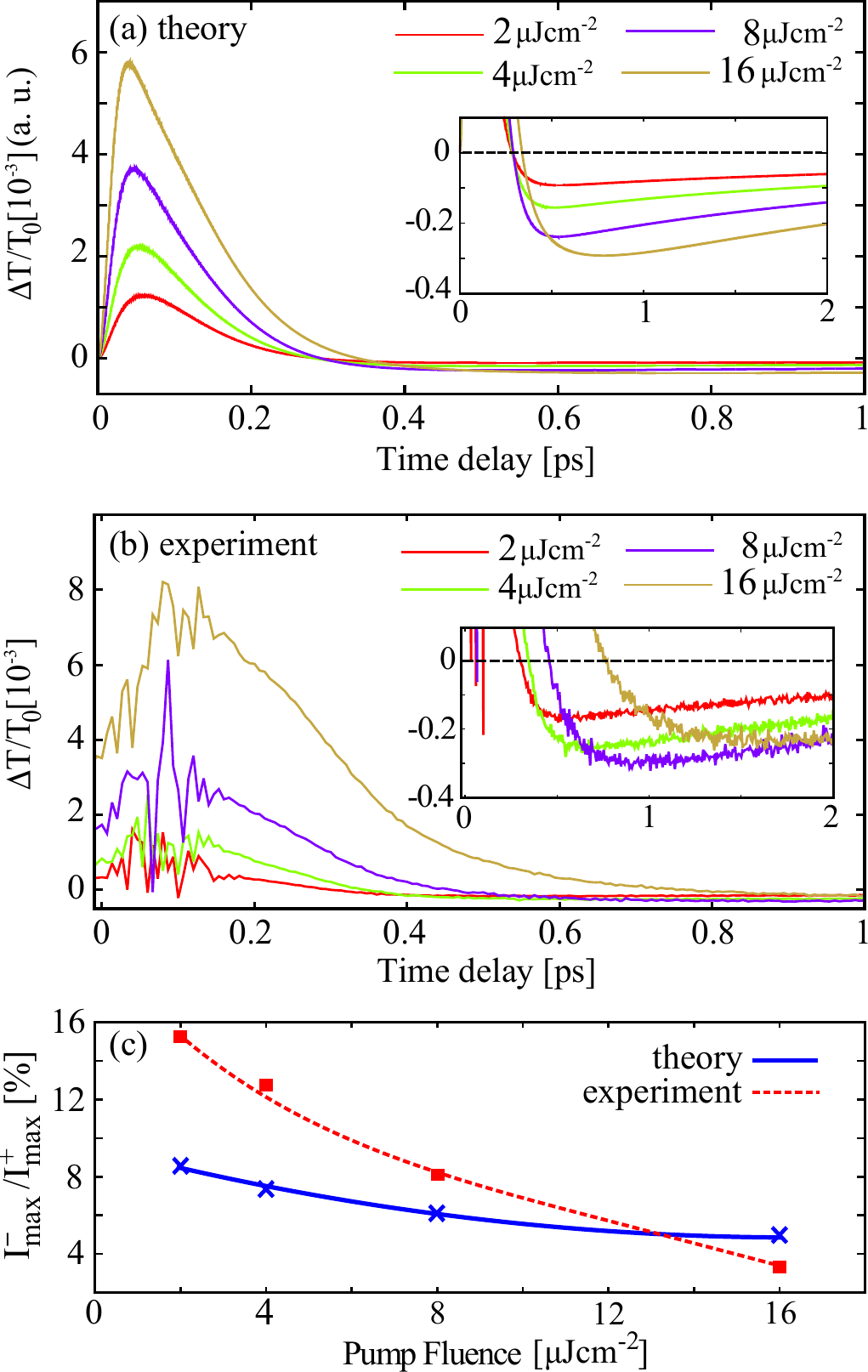}
  \end{center}
  \caption{ (a) Transient differential transmission spectrum for different pump fluences including the phonon-assisted intraband absorption processes.
	(b) Experimentally measured transient differential transmission at the same conditions (excitation energy of 1.5 eV and a pulse width of 50 fs).
	(c) Fluence-dependent ratio $\text{I$^-$}_{\text{max}}/\text{I$^+$}_{\text{max}}$ denoting the  maximal 
	negative and positive transient DT signal, respectively. The strongest relative impact of phonon-assisted intraband processes is predicted at small fluences - in good agreement with the experiment.
	}\label{fig3} 
\end{figure}

Figure \ref{fig3}(a) illustrates the impact of the phonon-assisted intraband absorption on the transient DT at different pump fluences covering the range of $2 - 24$ $\mu$J/cm$^2$.
The fast decay time $\tau_1$ exhibits a weak increase from $80$ to $110$ fs with the increasing fluence. In the case of weak excitation, the thermalized carrier distribution differs only slightly from the initial thermal distribution.
Due to the low population, the impact of Pauli blocking above the thermal tail is almost negligible with the consequence that the out-scattering from the excited states becomes very efficient and leads to a fast decay $\tau_1$. 
For stronger excitation, the efficient Pauli blocking suppresses the out-scattering explaining the increased $\tau_1$ time.\cite{winzer13-1}  
In contrast to the fast decay component, the slower $\tau_2$ time decreases approximately linearly from $3.3$ to $2.7$ ps with increasing pump fluence:
At high excitation, there is an increased number of excited carriers at the Dirac-point resulting in faster recombination of the carrier system.

To test the predicted behavior, we compare our results with recently performed experimental data [\onlinecite{winnerl13}] depicted in  Fig. \ref{fig3}(b).
Investigating increasing pump intensity, we find both in theory and experiment (i) similar timescales $\tau$, (ii) the same trend of the zero-crossing, and (iii) height trend as a function of pump intensity.
The absolute value of the negative transient DT signal increases with the pump fluence, cp. the insets of Fig. \ref{fig3}(a) and (b).
At high fluences, the intraband absorption is enhanced due to the larger number of available carriers in the conduction band. At the same time, increasing the fluence gives rise to a more efficient Pauli blocking, which increases the 
absorption bleaching and leads to a stronger positive DT signals.
Figure \ref{fig3}(c) shows the fluence-dependence of the ratio $\text{I$^-$}_{\text{max}}/\text{I$^+$}_{\text{max}}$ 
between  the  maximal negative ($\text{I$^-$}_{\text{max}}$) and positive ($\text{I$^+$}_{\text{max}}$) transient DT signal.
We find a good agreement between the theoretically predicted and experimentally measured behavior: The ratio clearly decreases with the fluence with a maximal ratios in the range of $10\%$  found in the weak excitation regime. 
The decrease can be traced back to the predominant role of the Pauli blocking at high fluences prevailing over the increased efficiency of intraband absorption. This also explains the observation both in theory and experiment that 
the zero crossing occurs at larger delay times with the increasing fluence, cf. the insets of Fig. \ref{fig3}(a) and (b).
The theoretically predicted values for $\tau_2$  correspond well with the timescale experimentally observed.

In conclusion, we propose a microscopic mechanism explaining the occurrence of the  transient negative differential transmission in graphene that has been observed
in several recent pump-probe experiments. In agreement with experimental results, our calculations reveal that the interplay of interband and phonon-assisted intraband absorption processes provides a qualitative 
explanation of this effect: phonon-induced processes open an additional absorption channel assisted by simultaneous absorption or emission of phonons that are 
required to fulfill the conservation of momentum.

We acknowledge the financial support from the Deutsche Forschungsgemeinschaft (DFG)
through SFB 658 (A. K., F. K.) and SPP 1459 (E. M., S. W.). E. M. is also thankful to the Einstein Foundation Berlin. 
T. W. was supported by the third Stage of Brain Korea 21 Plus Project
Division of Creative Materials in 2014.

\end{document}